\documentclass[twocolumn,showpacs,preprintnumbers,amsmath,amssymb]{revtex4}
\input psfig.sty

\usepackage{graphicx}
\usepackage{dcolumn}
\usepackage{bm}

\begin{document}



\title{Cosmological consequences of a Chaplygin gas dark energy}

\author{Abha Dev} \email{abha@ducos.ernet.in}  
\affiliation{%
Department of Physics and Astrophysics, University of Delhi,
Delhi-110007, India  
}%

\author{J. S. Alcaniz} \email{alcaniz@astro.washington.edu}
\affiliation{Astronomy Department, University of Washington, Seattle,
Washington, 98195-1580, USA
}%

\author{Deepak Jain} \email{deepak@ducos.ernet.in} 
\affiliation{%
Deen Dayal Upadhyaya College, University of Delhi,
Delhi - 110015, India  
}%

\date{\today}

\begin{abstract}
A combination of recent observational results has given rise to what is currently known as the dark energy
problem. Although several possible candidates have been extensively discussed in the literature to date the 
nature of this dark energy component is not well understood at present. In
this paper we investigate some cosmological implications of another dark energy candidate: an exotic fluid
known as the Chaplygin gas, which is characterized by an equation of state $p = -A/\rho$, where $A$ is a
positive
constant. By assuming a flat scenario driven by non-relativistic matter plus a Chaplygin gas dark energy we
study the influence of such a component on the statistical properties of gravitational lenses. A comparison
between
the predicted age of the universe and the latest age estimates of globular clusters is also included and the
results briefly discussed. In
general, we find that the behavior
of this class of models may be interpreted as an intermediary case between the standard and $\Lambda$CDM
scenarios.

\end{abstract}

\pacs{98.80.Es; 95.35.+d; 98.62.Sb}
\maketitle

\section{Introduction}

From a large number of observational evidence, the currently favoured cosmological model is flat, accelerated 
and composed of $\sim 1/3$ of matter (barionic + dark) and $\sim 2/3$ of a negative-pressure dark component,
usually named dark energy or ``quintessence". The nature of such an unclustered dark energy component,
however, is not very well understood at present, giving rise to many theoretical speculations. 

Certainly, the most extensively studied explanation for this dark energy problem is the vaccum
energy density or
cosmological
constant ($\Lambda$) although other interesting possibilities are also alive in the current literature. Some 
examples are: a very light scalar field $\phi$, whose effective potential $V(\phi)$ leads to 
an accelerated phase at the late stages of the Universe \cite{peebles}, a X-matter component \cite{turner},
which is simply characterized by an equation of state $p_x = \omega_x\rho_x$, where $-1 \leq \omega_x < 0$ 
and
that includes, as a particular case, models with a cosmological constant ($\Lambda$CDM), a vaccum decaying
energy density or a time varying $\Lambda$-term whose the present value of the cosmological constant
($\Lambda_o$) is a remnant of the primordial inflationary/deflationary stage \cite{ozer}, geometrical effects
from extra dimensions \cite{dvali} or still an exotic fluid, the so-called Chaplygin gas, whose equation of
state is given by 
\begin{equation}
p = -A/\rho,
\end{equation}
where $A$ is a positive constant \cite{kamen}.

All the above mentioned candidates for quintessence have interesting features that make them at some level
compatible with the recent obervational facts (see, for example, \cite{tur,cald,cunha,jailson}). Although most
of these
scenarios have been extensively explored in the recent literature, in the case of a Chaplygin gas-type dark
energy, however, only few analysis have focused attention on its cosmological consequences. From a theoretical
viewpoint, an interesting connection between the Chaplygin gas equation of state and String theory has been 
identified \cite{bord,jac,bilic}. As explained in \cite{fabris,bento}, a Chaplygin gas-type equation of
state is associated with the parametrization invariant Nambu-Goto $d$-brane action in a $d + 2$
spacetime. In the light-cone parametrization, such an action reduces itself to the action of a Newtonian fluid
which obeys Eq. (1) so that the Chaplygin gas corresponds effectively to a gas of $d$-branes in a $d + 2$
spacetime. 
Moreover, the Chaplygin gas is the only gas known to admit supersymmetric
generalization \cite{jac}. 

From the observational viewpoint, it has been argued that the Chaplygin gas may unify the
cold dark
matter and the dark energy scenarios \cite{bilic}. The reason for such a belief is the general behaviour of
the Chaplygin gas equation of state: it can behave as cold dark matter at small scales and as a
negative-pressure dark energy
component at large scales. Recently, Fabris {\it et al.} \cite{fabris1} analysed a
cold dark matter plus a Chaplygin gas scenario in the light of type Ia supernovae data
(SNe Ia). As a general
result, they found a universe completely dominated
by the Chaplygin gas as the best fit model. More recently, Avelino {\it et al.} \cite{avelino} used a larger
sample of SNe Ia and the shape of the matter power spectrum to show that such data restrict the model to a
behaviour that closely matches that of a $\Lambda$CDM models while Bento {\it et al.} \cite{bento1} showed
that the location of the CMB peaks imposes tight constraints on the free parameters of the model.

The aim of this paper is to explore some other observational consequences of a Chaplygin gas 
dark energy. We mainly focus our attention on the constraints from statistical properties
of gravitationally
lensed quasars (QSOs) on the Eq. (1). We also investigate other observational quantities like the deceleration
parameter, the acceleration redshift and the
expanding age of the the universe. To obtain such results we assume a flat model driven by non-relativistic
matter plus a Chaplygin gas dark energy component (from now on CgCDM). 

This paper is organized in the following way. 
In Sec. II the field equations and distance formulas are presented. We also derive the expression for the
deceleration parameter and discuss the redshift at which the accelerated expansion
begins. The predicted age of the Universe in the context of CgCDM models is briefly discussed in Sec. III. We
then proceed to analyse the constraints from lensing statistics on these scenarios in  In Sec. IV.
We end the paper by summarizing the main results in the conclusion section.

\section{Field equations, deceleration parameter and distance Formulas}

Let us now consider the Friedmann-Robertson-Walker (FRW) line element ($c = 1$)
\begin{equation}
ds^2 = dt^2 - R^{2}(t) \left[{dr^{2} \over 1 - kr^{2}} + r^{2} (d
\theta^2 + \rm{sin}^{2} \theta d \phi^{2})\right],
\end{equation}
where $k = 0$, $\pm 1$ is the curvature parameter of the spatial section, $r$, $\theta$, and $\phi$ 
are dimensionless comoving coordinates, and $R(t)$ is the scale factor. Since the two components
(nonrelativistic matter and Chaplygin gas) are separately conserved, we use the energy conservation law
together with Eq. (1) to find the following expression for the Chaplygin gas density
\begin{equation} \label{limB}
\rho_{Cg} = \sqrt{A + B\left(\frac{R_o}{R}\right)^{6}},
\end{equation}
or, equivalently, 
\begin{equation}
\rho_{Cg} = \rho_{Cg_{o}}\sqrt{A_s + (1 - A_s)\left(\frac{R_o}{R}\right)^{6}},
\end{equation}
where the subscript $o$ denotes present day quantities, $B = \rho_{Cg_{o}}^{2} - A$ and $A_s =
A/\rho_{Cg_{o}}^{2}$ is a quantity related with the sound speed for the Chaplygin gas today. As can be seen
from Eq. (\ref{limB}), the Chaplygin gas interpolates between non-relativistic matter ($\rho_{Cg}(R
\rightarrow 0) \simeq \sqrt{B}/R^{3}$) and negative-pressure dark component regimes ($\rho_{Cg}(R \rightarrow
\infty)
\simeq \sqrt{A}$).

The Friedmann's equation for the kind of models we are considering is
\begin{eqnarray}
\frac{\dot R}{R} =  H_o  \left[\Omega_{\rm{m}}(\frac{R_o}{R})^{3} + \Omega_{Cg}\sqrt{A_s + (1 - A_s)
(\frac{R_o}{R})^{6}}\right]^{\frac{1}{2}}.
\end{eqnarray} 
In the above equation, an overdot denotes derivative with respect to time, $H_o =
100h$ ${\rm{km.s^{-1}Mpc^{-1}}}$ is the present day value of the Hubble parameter, and $\Omega_{\rm{m}}$ and
$\Omega_{Cg}$ are, respectively, the matter and the Chaplygin gas density parameters.

The deceleration parameter, usually defined as $q_o = -R\ddot{R}/\dot{R}^{2}|_{t_{o}}$, now takes 
the following form
\begin{equation}
q_o = \frac{\frac{3}{2}\left[\Omega_{\rm{m}} + \Omega_{Cg}(1 - A_s)\right]}{\Omega_{\rm{m}} + \Omega_{Cg}}
- 1.
\end{equation}
As one may check, for $A_s = 0$ and $A_s = 1$, the above expressions reduce to the standard and
$\Lambda$CDM models, respectively.

\begin{figure}
\vspace{.2in}
\centerline{\psfig{figure=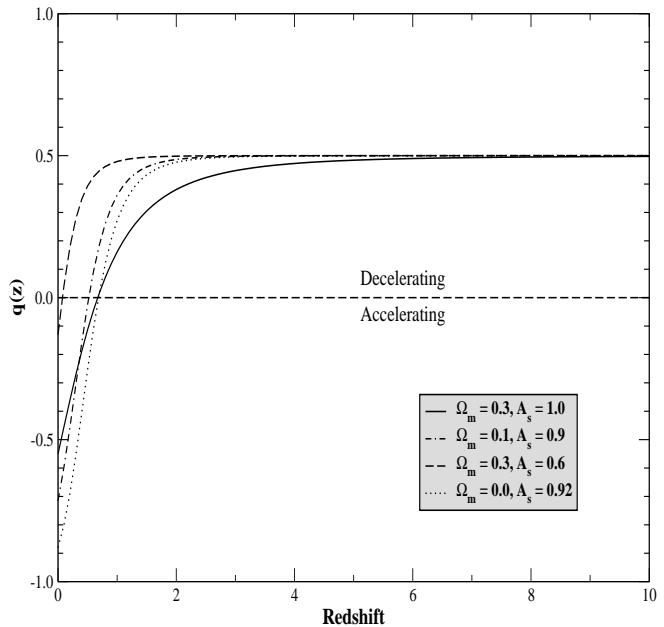,width=3.5truein,height=3.5truein,angle=-90}
\hskip 0.1in} 
\caption{Deceleration parameter as a function of redshift for some selected values of 
$\Omega_{\rm{m}}$ and $A_s$. The horizontal line labeled decelerating/accelerating ($q_o = 0$) divides models 
with a decelerating or accelerating expansion at a given redshift.}
\end{figure}

Figure 1 shows the behavior of the deceleration parameter as a function of redshift for selected 
values of $\Omega_{\rm{m}}$ and $A_s$.  The best fit $\Lambda$CDM case is also showed for the sake of 
comparison ($A_s = 1$). Note that the value of $A_s$ determines the acceleration redshift $z_a$. At late
times, a CgCDM
model with $\Omega_{\rm{m}} = 0.1$ and $A_s = 0.9$ accelerates faster than a $\Lambda$CDM scenario with
$\Omega_{\rm{m}} = 0.3$. In such a model the accelerated expansion begins at $z_a \simeq 0.51$. For the 
best fit model found in Ref. \cite{fabris}, i.e., $\Omega_{\rm{m}} = 0$ and $A_s = 0.92$, the universe is
strongly accelerated today with the accelerated phase begining at $z_a \simeq 0.68$ whereas for $A_s =1$ and
$A_s
= 0.6$ and $\Omega_{\rm{m}} = 0.3$ we find, respectively, $z_a \simeq 0.67$ and $z_a \simeq 0.07$.

\begin{figure}
\vspace{.2in}
\centerline{\psfig{figure=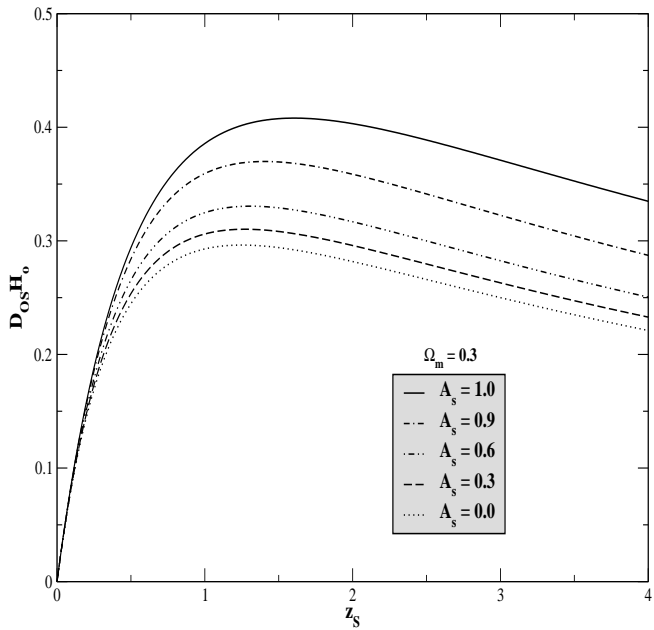,width=3.5truein,height=3.5truein,angle=-90}
\hskip 0.1in} 
\caption{Dimensionless angular diameter distance as a function of the source redshift ($z_S$) for some
selected values of $A_s$. In all curves the value of the matter density parameter has been fixed
($\Omega_{\rm{m}} = 0.3$).}
\end{figure}

From Eqs. (2) and (5), it is straightforward to show that the comoving distance $r_1(z)$ to a light source
located at $r = r_1$ and $t = t_1$ and observed at $r = 0$ and $t = t_o$ can be written as
\begin{equation}  
r_1(z) = \frac{1}{R_o H_o}\int_{x'}^{1} {dx \over x^{2}f(x, \Omega_{\rm{m}}, A_s)}, 
\end{equation} 
where  $x' = {R(t) \over R_o} = (1 + z)^{-1}$ is a convenient integration variable and the dimensionless
function $f(x, \Omega_{\rm{m}}, A_s)$ is given by 
\begin{eqnarray}
f(x, \Omega_{\rm{m}}, A_s)  =  \left[\frac{\Omega_{\rm{m}}}{x^{3}}  + (1 - \Omega_{\rm{m}})\sqrt{A_s +
\frac{(1 - A_s)}{x^{6}}}\right]^{\frac{1}{2}}.
\end{eqnarray} 

In order to derive the constraints from lensing statistics in Sec. IV we shall deal with the concept
of angular diameter distance. For the class of models here investigated, the angular diameter distance,
$D_{LS}(z_L, z_S) = {R_or_1(z_L, z_S) \over (1 + z_S)}$, between two objects,
for example a lens at $z_L$ and a source (galaxy) at $z_S$, reads  
\begin{eqnarray}
D_{LS}(z_L, z_S) & = & \frac{H_o^{-1}}{(1 + z_S)} \times \\ \nonumber & & \times  
 \int_{x'_S}^{x'_L} {dx \over x^{2} f(x, \Omega_{\rm{m}}, A_s)} .
\end{eqnarray}

In Fig. 2 we show the dimensionless angular diameter distance between an observer $O$ and the source $S$
($D_{OS}H_o$) as a function of the source redshift ($z_S$) for $\Omega_{\rm{m}} = 0.3$ and selected values of
$A_s$. As physically expected, the larger the value of $A_s$ the larger the distance that is predicted between
two redshifts. This result shows that, for the value of $\Omega_{\rm{m}}$ considered, the behaviour of this
class of CgCDM models may be interpreted as an
intermediary case between the $\Lambda$CDM ($A_s = 1$) and the Einstein-de Sitter ($A_s = 0$) scenarios. This
particular feature of CgCDM models may be important for the lensing statistics analysis because, as is well
known, the large distances predicted by $\Lambda$CDM models make the lensing constraints on the
vaccum energy contribution very restrictive (see, for instance, \cite{1CSK}). In this
concern, we expect that the constraints from this particular test will be weaker for CgCDM scenarios than for
their $\Lambda$CDM counterparts. It is worth mentioning that
the behavior of CgCDM cosmology can be very different from that one present by $\Lambda$CDM scenarios and
general quintessence cosmologies. For example, as shown in Ref. \cite{gorini}, the trajectories of the
statefinder parameters \cite{star} in CgCDM scenarios differs considerably 
from the one presented by quintesence or $\Lambda$CDM models. As commented in \cite{gorini}, the statefinder
diagnostic combined with future supernovae observations (as, for example,
the SNAP mission) may be able to discriminate
between CgCDM and general quintessence cosmologies. More recently, an analysis for the location of the CMB
peaks showed that CgCDM models and $\Lambda$CDM have very different predictions for large values of the
parameter $A_s$ \cite{bento1}.

\section{The Age of the Universe}

The predicted age of the Universe for the class of CgCDM models considered in this paper is given by
\begin{equation}  
t_o = \frac{1}{H_o}\int^{1}_{0} {dx \over xf(x, \Omega_{\rm{m}}, A_s)}.
\end{equation}
As widely known, a
lower bound for this quantity can be estimated in a variety of
different ways. For instance, Oswalt {\it et al.} \cite{osw},
analyzing the cooling sequence of white dwarf stars found a lower
age limit for the galactic disk of 9.5 Gyr. Later on, a value of
15.2 $\pm$ 3.7 Gyr was also determined using radioactive dating of
thorium and europium abundances in stars \cite{cowan}. In this
connection, the recent age estimate of an extremely metal-poor
star in the halo of our Galaxy (based on the detection of the
385.957 nm line of singly ionized $^{238}$U) indicated an age of
12.5 $\pm$ 3 Gyr \cite{cay}. Another important way of estimating a lower limit to
the age of the Universe is dating the oldest stars in globular clusters. Such
estimates, however, have oscillated considerably since the
publication of the statistical parallax measures done by
Hipparcos. Initially, some studies implied in a lower limit of 9.5
Gyr at $95\%$ confidence level (c.l.) \cite{chab}.
Nevertheless, subsequent studies \cite{chab1} using new
statistical parallax measures and updating some stellar model
parameters, found 13.2 Gyr with a lower limit of 11 Gyr at 95$\%$
c.l., as a corrected mean value for age estimates of globular clusters (see also
\cite{carreta}). Such a value implies that the Einstein-de Sitter model is ruled out
for $h \geq 0.50$, while the most recent measurements of $h$ point
consistently to $h \geq 0.65$ \cite{gio,f1}. These results
are also in accordance with recent estimates based on rather
different methods for which the ages of the oldest globular clusters in our
Galaxy fall on the interval 13.8 - 16.3 Gyr \cite{rengel}.

\begin{figure}
\vspace{.2in}
\centerline{\psfig{figure=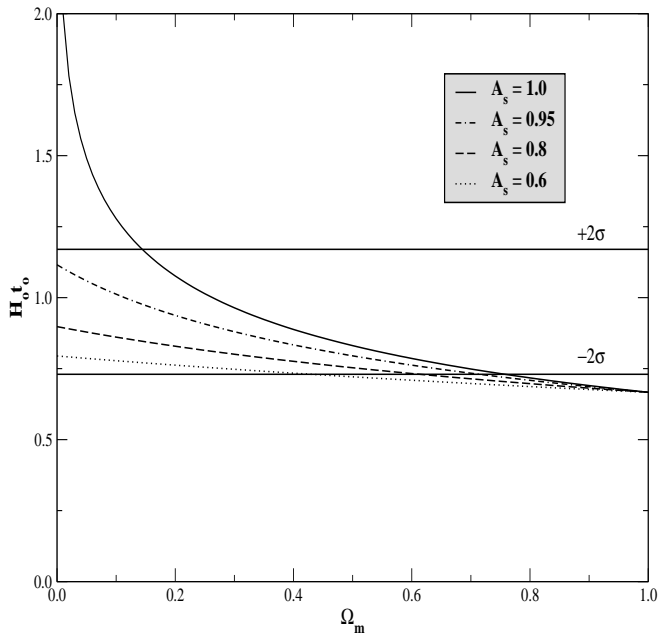,width=3.5truein,height=3.5truein,angle=-90}
\hskip 0.1in} 
\caption{$H_ot_o$ as a function of the matter density parameter for some values of $A_s$. Horizontal lines
correspond to $\pm 2\sigma$ limits of the age parameter $H_ot_o = 0.95 \pm 0.11$.}
\end{figure}

By assuming $t_o = 13 \pm 1$ Gyr as a median value for the most recent age estimates of globular clusters and 
$H_o = 72 \pm 8$ ${\rm{Km.s^{-1}.Mpc^{-1}}}$, in accordance with the final results of the {\it Hubble Space
Telescope} Key Project \cite{f1}, we find $H_ot_o = 0.95 \pm 0.11$, a value that is compatible with the
estimates discussed above, as well as very close to some determinations based on SNe Ia data
\cite{riess,tonry}. In Fig. 3 we show the dimensionless age parameter $H_ot_o$ as a function of
$\Omega_{\rm{m}}$ for some selected values of $A_s$. Horizontal dashed lines indicate $\pm 2\sigma$ of the age
parameter for the values of $H_o$ and $t_o$ considered here. Similarly to the discussion for the angular
diameter distance, for a fixed value of $\Omega_{\rm{m}}$ the predicted age of the Universe is larger for
larger values of $A_s$. If $\Omega_{\rm{m}} = 0.2 - 0.4$, as sugested by dynamical estimates on scales up to
about $2h^{-1}$ Mpc \cite{calb}, we find $A_s \geq 0.96$ (see also \cite{jailn}).

\section{Constraints from Lensing Statistics}
 
In order to constrain the parameters $\Omega_{\rm{m}}$ and $A_s$ from lensing statistics we work with a sample
of 867 ($z > 1$) high
luminosity  optical quasars which includes 5 lensed quasars. 
This sample consists of data from the following  optical lens surveys:
HST Snapshot survey \cite{HST}, Crampton survey 
\cite{Crampton}, Yee survey \cite{Yee},
Surdej survey \cite{Surdej}, NOT Survey \cite{Jaunsen} 
and FKS survey \cite{FKS}.

The differential probability $d\tau$ of a beam having a lensing 
event in traversing $dz_{L}$ is \cite{TOG,FFKT}
\begin{equation} \label{20}
d\tau = F^{*}(1 +
z_{L})^{3}\left({D_{OL}D_{LS}\over R_0 D_{OS}}\right)^{2}
{1\over R_0}{dt \over dz_{L}} dz_{L},
\end{equation}
where 
\begin{equation} {dt\over dz_L} = {H_o^{-1} \over (1 + z_L) f(x, \Omega_{\rm{m}}, A_s)},
\end{equation}
and
\begin{equation}
F^* = {16\pi^{3}\over{c
H_{0}^{3}}}\phi_\ast v_\ast^{4}\Gamma\left(\alpha + {4\over\gamma} +1\right).
\end{equation}
In Eq. (\ref{20}), $D_{OL}$, $D_{OS}$ and $D_{LS}$ are, respectively, 
the angular diameter distances from the observer to the lens, from the
observer to the source and between the lens and the source. 
We use the Schechter luminosity function with the lens parameters 
for E/SO galaxies taken from 
Madgwick {\it et al.} \cite{Mad}, i.e., $\phi_{\ast} = 0.27 h^{3}\,10^{-2} 
{\rm{Mpc}}^{-3}$, $\alpha = -0.5$, $\gamma = 4$, $v_{\ast} = 220\,{\rm{km/s}}$ and
$F^{*} = 0.01$. 

\begin{figure}
\vspace{.2in}
\centerline{\psfig{figure=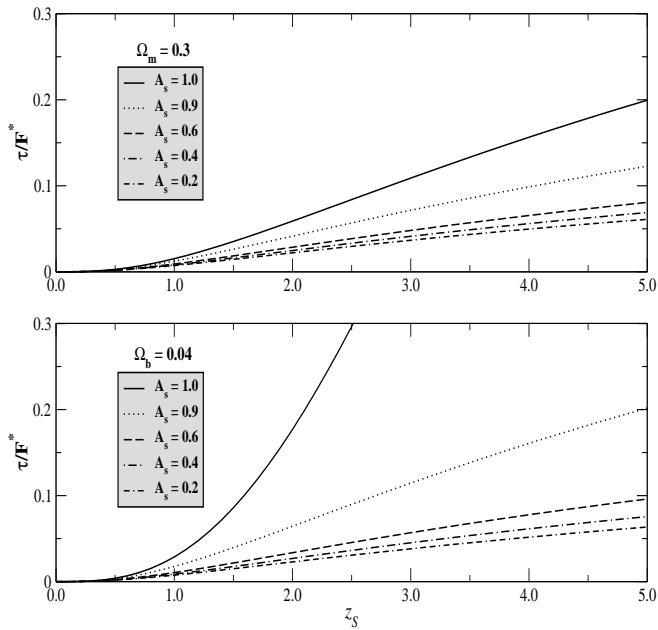,width=3.5truein,height=3.5truein,angle=-90}
\hskip 0.1in} 
\caption{The normalized optical depth ($\tau/F^{*}$) as a function of the source redshift ($z_S$) for some
selected  values of $A_s$. Upper panel: CgCDM models with $\Omega_{\rm{m}} = 0.3$. Lower panel: a universe
dominated by baryonic
matter ($\Omega_{b} = 0.04$) and the Chaplygin gas.}
\end{figure}

The total optical depth is obtained by integrating $d\tau$ along the
line of sight from $z_O$ ($z = 0$) to $z_S$. One obtains
\begin{equation}
\tau(z_S) = \frac{F^{*}}{30} \left[D_{OS}(1+z_{L})\right]^{3}\, R_{o}^{3}.
\label{atau}
\end{equation}
In Fig. 4 we show the normalized optical depth as a function of the source redshift ($z_S$) for values of $A_s
=$ 0.2, 0.4, 0.6, 0.9 and 1.0. 
Two different cases are illustrated: a conventional CgCDM model
with the matter density parameter fixed at $\Omega_{\rm{m}} = 0.3$ and 
a Chaplygin gas + baryonic matter model
with $\Omega_{b} = 1 - \Omega_{Cg} = 0.04$. As discussed earlier, the reason for considering the later case is
because one of the strongest claims for a Chaplygin gas dark energy is the possibility of a unified
explanation for the dark matter and dark energy problems \cite{bilic}. In this case, one might expect that the
only two components of the Universe would be the Chaplygin gas and the baryonic matter. Note that in both
cases an increase in $A_s$ at fixed $\Omega_{\rm{m}}$ tends to increase the optical depth for lensing. For
example, for $\Omega_{\rm{m}} = 0.3$, the value of $\tau/F^{*}$ for $A_s = 0.2$ at $z_S = 3.0$ is down from
the 
$\Lambda$CDM ($A_s = 1$) value by a factor of $\sim$ 2.97, while at the same redshift, 
$\tau/F^{*}$ for $A_s = 0.4$ is down from that for $A_s = 1.0$ by $\sim$ 2.63. By fixing the value of $A_s$,
for example, $A_s = 0.6$, we observe that the value of $\tau/F^{*}$ is smaller for a universe with
$\Omega_{\rm{m}} = 0.3$ than for a universe composed only of the Chaplygin gas + baryonic matter ($\Omega_{b}
= 0.04$) by a factor of $\sim$ 1.18. This 
increase of the optical depth as the value of $A_s$ is increased (at a fixed 
$z_S$  and $\Omega_{\rm{m}}$) is an expected consequence since this model more closely approaching the
$\Lambda$CDM case as $A_S \rightarrow 1$.

The likelihood function is defined by
\begin{equation} 
{\cal{L}} = \prod_{i=1}^{N_{U}}(1-p^{'}_{i})\,\prod_{k=1}^{N_{L}}
p_{k}^{'}\,p_{ck}^{'},
\label{LLF}
\end{equation}
where $N_{L}$ is the number of multiple-imaged lensed quasars, $N_{U}$ is the number of unlensed 
quasars, and $p_{k}^{'}$ and $p_{ck}^{i}$ are, respectively, the probability of quasar $k$ to be lensed and
the 
configuration probability. These quantities are defined by 
\begin{equation}
p^{'}_{i}(m,z) = p_{i}\int \frac{ d(\Delta\theta)\,
p_{c}(\Delta\theta)\emph{B}(m,z,M_{f}(\Delta\theta),M_{2})}{\emph{B}(m,z,M_{0},M_{2})}
\label{prob2}
\end{equation}
and
\begin{equation}
p^{'}_{ci} =
p_{ci}(\Delta\theta)\,\frac{p_{i}}{p_{i}^{'}}\,\frac{\emph{B}(m,z,M_{f}(\Delta
\theta),M_{2})} 
{\emph{B}(m,z,M_{0},M_{2})},
\label{confi}
\end{equation}
where
\begin{equation}
p_{c}(\Delta\theta) =
\frac{1}{\tau(z_{S})}\,\int_{0}^{z_{S}}\frac{d^{2}\tau}{dz_{L}d(\Delta\theta)}
\,dz_{L}
\label{pcphi}
\end{equation}
and 
\begin{equation}
M_{f} = M_{0}(f+1)/(f-1)\quad \mbox{with}\quad  f = 10^{0.4\,\Delta m(\theta)}.
\label{Mf}
\end{equation}

The magnification bias, ${\bf B}(m,z)$, is 
considered in order to take into account the increase in the
apparent brightness of a quasar due to lensing which, in turn, increases  the
expected number of lenses in flux limited sample.
The bias factor for a quasar at redshift $z$ with apparent magnitude
$m$ is  given by \cite{FFKT,1CSK}
\begin{equation}
{\bf B}(m,z) = M_{0}^{2}\, \emph{B}(m,z,M_{0},M_{2}),
\label{bias}
\end{equation} 
where
\begin{eqnarray}
\emph{B}(m,z,M_{1},M_{2})& =&
2\,\left(\frac{dN_{Q}}{dm}\right)^{-1}\int_{M_{1}}^{M_{2}}\frac{dM}{M^{3}}\, \\
&& \times \,\frac{dN_{Q}}{dm}(m+2.5\log(M),z). \nonumber
\label{bias1}
\end{eqnarray}
In the above equation $({dN_{Q}(m,z)}/{dm})$ is the measure of number of quasars with magnitudes 
in the interval $(m,m+dm)$ at redshift $z$. Since we are modeling the lens by a singular isothermal model 
profile, $M_0 = 
2$, we adopt $M_{2} = 10^{4}$ in the numerical computation.

For the quasar luminosity function we use Kochanek's ``best model'' \cite{1CSK}
\begin{equation}
\frac{dN_{Q}}{dm}(m,z) \propto
(10^{-a(m-\overline{m})}+10^{-b(m-\overline{m})})^{-1}
\label{lum},
\end{equation}
where
\begin{equation}
\overline{m} = \left\{ \begin{array}{ll}
                   m_{o}+(z-1)    & \mbox{for $z < 1$} \\
                   m_{o}          & \mbox{for $1 < z \leq 3$} \\
                   m_{o}-0.7(z-3) & \mbox{for $z > 3$}
                   \end{array}
               \right. \
\end{equation}
and we assume $a = 1.07 \pm 0.07$, $b = 0.27 \pm 0.07$ and $m_{o} =
18.92 \pm 0.16$ at B magnitude \cite{1CSK}.

Due to selection effects the survey can detect lenses with
magnification larger than a certain magnitude $M_f$  given by equation
(\ref{Mf}) which becomes the lower limit in equation (\ref{bias1}).
To obtain selection function corrected probabilities, we follow
\cite{1CSK} and divide our sample into two parts, namely, the ground based 
surveys and the HST survey.

\begin{figure}
\vspace{.2in}
\centerline{\psfig{figure=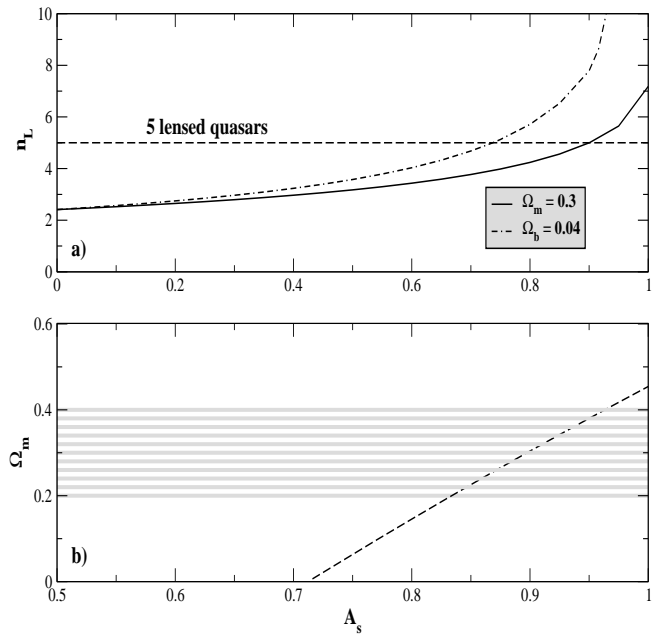,width=3.5truein,height=3.5truein,angle=-90}
\hskip 0.1in} 
\caption{{\bf{a)}} Predicted number of lensed quasars as a function of 
$A_s$ for $\Omega_{\rm{m}} = 0.3$ (solid line) and $\Omega_{b} = 0.04$ (dashed-dotted line) and 
image separation $\Delta \theta \leq 4$. {\bf{b)}} Contour for five lensed quasars in the 
parametric space $A_s - \Omega_{\rm{m}}$. The shadowed horizontal region corresponds to 
the observed range $\Omega_{\rm{m}} = 0.3 \pm 0.1$ \cite{calb}.}
\end{figure}

From Eq. (\ref{LLF}) we find that the maximum value of the likelihood function is located at 
$\Omega_{\rm{m}} = 0.4$ and $A_s = 1.0$. At the 1$\sigma$ level, however, almost the entire range of $A_s$ is
compatible with the observational data for values of $\Omega_{\rm{m}} = 0 - 1$. As observed earlier (see Sec.
II), this result suggests that a large class of CgCDM scenarios is in accordance with the current
gravitational
lensing data. For the sake of comparison, we also analyse some possible differences between our best-fit value
and the one obtained for general quintessence scenarios with an equation of state $p_x = \omega_x \rho_x$
(XCDM)
\cite{turner}. For example, for XCDM models a similar analysis shows that the maximum value of the likelihood
function is located at 
$\Omega_{\rm{m}} = 0.0$ and $\omega_x = -0.2$ \cite{waga}. Such a model corresponds to a decelerated universe
with a deceleration parameter $q_o = 0.2$ and a total expanding age of $8.1h^{-1}$ Gyr while our best-fit
CgCDM model corresponds to an accelerating scenario with $q_o = -0.39$ ($z_a = 0.44$) and a total age of the
order of $8.7h^{-1}$ Gyr. In Fig. 5a the expected number of lensed quasars, $n_L = \sum\, p_{i}^{'}$ (the
summation is
over a given
quasar sample), is displayed as a function of $A_s$. As indicated in the figure, the horizontal
dashed line indicates $n_L = 5$, that is the number of lensed quasars in our sample.  By this analysis, one
finds $A_s = 0.9$ ($\Omega_{\rm{m}} = 0.3$) and $A_s = 0.73$ ($\Omega_{b} = 0.04$). In Fig. 3b we show 
the contour for five lensed quasars in the parametric space $A_s - \Omega_{\rm{m}}$. The 
shadowed horizontal region corresponds to the observed range $\Omega_{\rm{m}} = 0.3 \pm 0.1$ 
\cite{calb}. As a general result, this analysis provides $\Omega_{\rm{m}} \leq 0.45$ and $A_s \geq 0.72$. We
also observe that the higher the value of $\Omega_{\rm{m}}$ the higher the value of $A_s$ that
is required to fit these data.

\section{Conclusion}

The search for alternative cosmologies is presently in vogue and the leitmotiv is the observational support
for an accelerated universe provided by the SNe Ia results. In general, such alternative scenarios contain an
unkown 
negative-pressure dark component that explains the SNe Ia results and reconciles the inflationary 
flatness prediction ($\Omega_{\rm{T}} = 1$) with the dynamical estimates of the quantity of matter 
in the Universe ($\Omega_{\rm{m}} \simeq 0.3 \pm 0.1$). In this paper we have focused our attention 
on another dark energy candidate: the Chaplygin gas. We showed that the predicted age of the Universe in the
context of CgCDM models is compatible with the most recent age estimates 
of globular clusters for values of $\Omega_{\rm{m}} \simeq 0.2$ and $A_s \geq 0.96$. We also studied the
influence of such a component on the statistical properties of gravitational lensing. At $1\sigma$ level we
found that a large class of these scenarios is in agreement with the current lensing data with the maximum of
the likelihood function (Eq. \ref{LLF}) located at $\Omega_{\rm{m}} = 0.4$ and $A_s = 1.0$. As a general
result, the predicted number of lensed quasars requires $\Omega_{\rm{m}} \leq 0.45$ and $A_s \geq 0.72$.

\begin{acknowledgments}
The authors are very grateful to Raimundo Silva Jr. for helpful discussions
and a critical reading of the manuscript. JSA is supported by the Conselho Nacional de Desenvolvimento 
Cient\'{\i}fico e Tecnol\'{o}gico (CNPq - Brasil) and CNPq (62.0053/01-1-PADCT III/Milenio).
\end{acknowledgments}


\end{document}